\documentclass{amsart}
\usepackage{amssymb}

\usepackage{graphicx}

\input{tcilatex}
\input tcilatex

\begin{document}
\title[Information, covariances, evolutionary genetics]{Information and (co-)variances in discrete evolutionary genetics involving
solely selection}
\author{Thierry E. Huillet}
\address{Laboratoire de Physique Th\'{e}orique et Mod\'{e}lisation\\
CNRS-UMR 8089 et Universit\'{e} de Cergy-Pontoise\\
2 Avenue Adolphe Chauvin, F-95302, Cergy-Pontoise, France\\
Thierry.Huillet@u-cergy.fr}
\maketitle

\begin{abstract}
The purpose of this Note is twofold: First, we introduce the general
formalism of evolutionary genetics dynamics involving fitnesses, under both
the deterministic and stochastic setups, and chiefly in discrete-time. In
the process, we particularize it to a one-parameter model where only a
selection parameter is unknown. Then and in a parallel manner, we discuss the
estimation problems of the selection parameter based on a single-generation
frequency distribution shift under both deterministic and stochastic
evolutionary dynamics. In the stochastics, we consider both the celebrated
Wright-Fisher and Moran models.
\end{abstract}

\noindent \textbf{Keywords}: Evolutionary genetics, covariances, fitness
landscape, selection.\newline

\noindent \textbf{Topics: }Evolutionary processes (theory), Population
dynamics (Theory).\newline

\section{Introduction and outline}

In this Note, we revisit the basics of both the deterministic and stochastic
dynamics arising in discrete-time evolutionary genetics (EG). We start with
the haploid case with $K$ alleles before switching to the more tricky
diploid case. In the course of the exposition, we shall focus on a
particular one-parameter selection instance of the general fitness model for
which only the selection parameter is assumed to be unknown.\newline

Let us summarize and comment the content of Section $2$. In the
deterministic haploid case, the updates of the allele frequency
distributions are driven by the relative fitnesses of the alleles, ending up
in a state where only the fittest will survive. From the dynamics, it
appears that the mean fitness increases as time passes by, the rate of
increase being the variance in relative fitness. This constitutes the core
of the Fisher theorem of natural selection (FTNS). Introducing a discrete
version of the Fisher information on time brought about by the allelic
frequencies, it follows that one can identify this Fisher information with
the variance in relative fitness.

In the deterministic diploid case, there is a similar updating dynamics but
now on the full array of the genotype frequencies. When mating is random so
that the Hardy-Weinberg law applies, we may look at the induced marginal
allelic frequencies dynamics. It follows that the induced closed-form
allelic updating dynamics looks quite similar to the one occurring in the
haploid case except that the mean fitness now is a quadratic form in the
current frequencies whereas marginal fitnesses no longer are constant but
affine functions in these frequencies. In this context, the FTNS still holds
true but, as a result of the fitness landscape being more complex, there is
a possibility for a polymorphic equilibrium state to emerge. A short
incursion in the continuous-time setting shows that here again one can
identify the variance in relative fitness to a familiar Fisher information
on time brought about by the frequencies. If one rather looks at the partial
rate of increase of the mean fitness, one can identify it with half the
allelic variance in relative fitness which is that component of the full
genotypic variance in relative fitness which can be explained additively by
the alleles constituting the genotypes, \cite{Ew}. The remaining interaction
part of this decomposition of the full genotypic variance can naturally be
attributed to the dominance relationships between the alleles. In the
interpretation of some authors, including W. J. Ewens (\cite{Ew}, p. $64-67$%
), this last property based on the partial rate of increase rather
constitutes the essence of the FTNS. Using this circle of ideas, it follows
that looking at the partial rate of increase of the mean fitness also makes
sense when dealing with the full array of the genotype frequencies, no
matter what form of mating is at stake. It also involves an allelic variance
in relative fitness. In each case, we keep looking at the incarnation of
these results when dealing with the one-parameter selection model. So far
the results introduced and discussed can be found to be classical, our own
contribution being perhaps to put things in order and fix the notations and
formalism in a clear way. An excellent introduction to these and related
problems can be found in \cite{K1}. We believe that the following
developments can be considered as being entirely new.

To end up with Section $2$, we discuss the estimation problem of the
selection parameter based on both the current and shifted allele frequencies
observations. Similarly, we discuss this problem when the observable is a
general scalar output of the current frequency distribution. When looking at
the updating of this output, we encounter a particular incarnation of the
Price equation, \cite{F1}.\newline

Section $3$ is devoted to the stochastic version of these considerations
when the transitions in the constitutive allelic population sizes are given
by a $K-$dimensional Wright-Fisher model with total constant-size (see \cite
{Ew} and \cite{Mar}). We show that the mean of the increment of the random
absolute mean fitness is positive, whereas its rate of increase differs from
its variance. We suggest that when the size of the total allelic population
goes to infinity, one should recover part of the marginal deterministic
theory. In the selection example, we compute the classical Fisher
information on the selection parameter and exhibit its possible use in the
estimation problem. We finally present some comparative issues pertaining to
a related model of fundamental importance in the context of stochastic EG:
the Moran model.

Lots remain to be done in the same spirit, in particular including mutations
and considering the multi-loci case with recombination. We again emphasize
that in our models, there are no mutations included.

\section{EG theory: the deterministic point of view}

We start with the haploid case before moving to the diploid case, see (\cite
{Ew} and \cite{K1} for similar concerns).

\subsection{Single locus: haploid population with $K$ alleles}

Consider $K$ alleles $A_{k}$, $k=1,...,K$ attached to a single locus.
Suppose the current time-$t$ allelic frequency distribution on the $K-$%
simplex $S_{K}$ is given by $x_{k}$, $k=1,...$,$K$. Let $\mathbf{x}:=x_{k}$, 
$k=1,...,K$ stand for the column-vector\footnote{%
In the sequel, a boldface variable, say $\mathbf{x}$, will represent a
column-vector so that its transpose, say $\mathbf{x}^{*}$, will be a
line-vector.} of these frequencies with $\left| \mathbf{x}\right|
:=\sum_{k}x_{k}=1$. Let $w_{k}>0$, $k=1,...,K$ be the absolute fitness of
allele $A_{k}.$ Let 
\begin{equation}
\overline{w}_{k}\left( \mathbf{x}\right) =\frac{w_{k}}{w\left( \mathbf{x}%
\right) }  \label{1}
\end{equation}
be the relative fitness of allele $A_{k}$ where $w\left( \mathbf{x}\right)
:=\sum_{l}w_{l}x_{l}$ represents the mean fitness of the population at time $%
t$. We shall also let 
\begin{equation}
\sigma ^{2}\left( \mathbf{x}\right) =\sum_{k=1}^{K}x_{k}\left( w_{k}-w\left( 
\mathbf{x}\right) \right) ^{2}  \label{2}
\end{equation}
stand for the variance in absolute fitness and 
\begin{equation}
\overline{\sigma }^{2}\left( \mathbf{x}\right) =\sum_{k=1}^{K}x_{k}\left( 
\overline{w}_{k}\left( \mathbf{x}\right) -1\right) ^{2}=\sigma ^{2}\left( 
\mathbf{x}\right) /w\left( \mathbf{x}\right) ^{2}  \label{3}
\end{equation}
will be the variance in relative fitness.

From the deterministic EG point of view, the discrete-time update of the
allele frequency distribution on the simplex $S_{K}$ is given by\footnote{%
The symbol $^{\prime}$ is a common and useful notation to denote the updated
frequency}

\begin{equation}
x_{k}^{\prime }=p_{k}\left( \mathbf{x}\right) \text{, }k=1,...,K.  \label{4}
\end{equation}
where $p_{k}\left( \mathbf{x}\right) :=x_{k}\overline{w}_{k}\left( \mathbf{x}%
\right) .$ The quantity $\overline{w}_{k}\left( \mathbf{x}\right) -1$
interprets as the frequency-dependent Malthus growth rate parameter of $%
x_{k}.$

The vector $\mathbf{p}\left( \mathbf{x}\right) :=p_{k}\left( \mathbf{x}%
\right) $, $k=1,...,K$, maps $S_{K}$ into $S_{K}$. In vector form, with 
\textbf{$\overline{\mathbf{w}}$}$\left( \mathbf{x}\right) :=\overline{w}%
_{k}\left( \mathbf{x}\right) $, $k=1,...,K$ and $D_{\mathbf{x}}:=$diag$%
\left( x_{k},k=1,...,K\right) $, the nonlinear deterministic EG dynamics
reads: 
\begin{equation*}
\mathbf{x}^{\prime }=\mathbf{p}\left( \mathbf{x}\right) =D_{\mathbf{x}}%
\overline{\mathbf{w}}\left( \mathbf{x}\right) =D_{\overline{\mathbf{w}}%
\left( \mathbf{x}\right) }\mathbf{x},
\end{equation*}
or, with $\Delta \mathbf{x}:=\mathbf{x}^{\prime }-\mathbf{x}$, the increment
of $\mathbf{x}$%
\begin{equation*}
\Delta \mathbf{x}=\left( D_{\overline{\mathbf{w}}\left( \mathbf{x}\right)
}-I\right) \mathbf{x.}
\end{equation*}

Avoiding the trivial case where fitnesses are all equal, without loss of
generality, we can assume that either $w_{1}\geq ...\geq w_{K}=1$ or $%
w_{1}\leq ...\leq w_{K}=1.$ Thus allele $A_{1}$ or $A_{K}$ has largest
fitness. The deterministic EG dynamics attains an equilibrium where only the
fittest will survive. The equilibrium is an extremal state of the boundary
of $S_{K}$.\newline

\textbf{Example (selection).} In general, all the $w_{k}$ are unknown but
sometimes, the set of unknowns can be reduced to $1$ as follows: Let $s>-1$
stand for a selection parameter. Let $a_{k}$, $k=1,...,K$ stand for a known $%
\left[ 0,1\right] -$valued decreasing sequence with $a_{1}=1,$ $a_{K}=0$ and
assume $w_{k}=1+sa_{k}$. The fitness landscape is $w\left( \mathbf{x}\right)
=1+sa\left( \mathbf{x}\right) $ where $a\left( \mathbf{x}\right)
:=\sum_{k}a_{k}x_{k}.$ A possible choice of $a_{k}$ is $a_{k}=\left(
K-k\right) /\left( K-1\right) $ leading to equally spaced fitnesses with $%
w_{k+1}-w_{k}=-s/\left( K-1\right) .$ An alternative choice is $a_{k}=\left(
K/k-1\right) /\left( K-1\right) $ with $w_{k+1}-w_{k}=-\frac{Ks}{K-1}\frac{1%
}{k\left( k+1\right) }.$ Depending on $s>0$ or $s<0$, the unit fitness $1$
is either the minimal or the maximal value of the ordered $w_{k}$s$.$
Although this particular model does not cover the class of all possible
fitnesses, its generality is sufficient for our purposes and allows a
considerable simplification of the exposition which otherwise would become
tedious. It does not alter the general line of thought in a major way. $%
\blacklozenge $\newline

According to the EG dynamical system (\ref{4}), for each $k$, the relative
fitness decreases as time passes by. Indeed, with $\Delta \overline{w}%
_{k}\left( \mathbf{x}\right) :=\overline{w}_{k}\left( \mathbf{x}^{\prime
}\right) -\overline{w}_{k}\left( \mathbf{x}\right) $%
\begin{equation*}
\Delta \overline{w}_{k}\left( \mathbf{x}\right) =\overline{w}_{k}\left( D_{%
\overline{\mathbf{w}}\left( \mathbf{x}\right) }\mathbf{x}\right) -\overline{w%
}_{k}\left( \mathbf{x}\right) =\frac{w_{k}}{w\left( D_{\overline{\mathbf{w}}%
\left( \mathbf{x}\right) }\mathbf{x}\right) }-\frac{w_{k}}{w\left( \mathbf{x}%
\right) }
\end{equation*}
\begin{equation*}
=\frac{w_{k}}{\sum_{l}w_{l}\overline{w}_{l}\left( \mathbf{x}\right) x_{l}}-%
\frac{w_{k}}{\sum_{l}w_{l}x_{l}}<0
\end{equation*}
because $\overline{w}_{l}\left( \mathbf{x}\right) =\frac{w_{l}}{w\left( 
\mathbf{x}\right) }$ and $w\left( \mathbf{x}\right)
^{2}<\sum_{l}w_{l}^{2}x_{l}$. However, unless the equilibrium state is
attained, the absolute mean fitness $w\left( \mathbf{x}\right) $ increases: 
\begin{equation*}
\Delta w\left( \mathbf{x}\right) =w\left( \mathbf{x}^{\prime }\right)
-w\left( \mathbf{x}\right) =\sum_{k}w_{k}\Delta x_{k}
\end{equation*}
\begin{equation*}
=\sum_{k}w_{k}x_{k}\left( \overline{w}_{k}\left( \mathbf{x}\right) -1\right)
=\frac{\sum_{k}w_{k}^{2}x_{k}}{w\left( \mathbf{x}\right) }-w\left( \mathbf{x}%
\right) >0.
\end{equation*}
The mean fitness is maximal at equilibrium. The rate of increase of $w\left( 
\mathbf{x}\right) $ is: 
\begin{equation}
\frac{\Delta w\left( \mathbf{x}\right) }{w\left( \mathbf{x}\right) }%
=\sum_{k}x_{k}\left( \overline{w}_{k}\left( \mathbf{x}\right) -1\right)
^{2}=\sum_{k}\frac{\left( \Delta x_{k}\right) ^{2}}{x_{k}}  \label{5}
\end{equation}
which is the variance in relative fitness $\overline{\sigma }^{2}\left( 
\mathbf{x}\right) $ defined in (\ref{3}). These last two facts are sometimes
termed the $1930$s Fisher fundamental theorem of natural selection (FTNS).%
\newline

\textbf{Remarks.}

$\left( i\right) $ The expression appearing in the right-hand side of (\ref
{5}) is also 
\begin{equation*}
\sum_{k}\frac{\left( \Delta x_{k}\right) ^{2}}{x_{k}}=\sum_{k}x_{k}\left( 
\frac{\Delta x_{k}}{x_{k}}\right) ^{2}.
\end{equation*}
The discrete frequency distribution $\mathbf{x}$ depends on the time
parameter $t\in \left\{ 0,1,2,...\right\} $ which is itself discrete. The
quantity $I_{\mathbf{x}}\left( t\right) :=\sum_{k}x_{k}\left( \frac{\Delta
x_{k}}{x_{k}}\right) ^{2}$ may therefore be interpreted as a discrete
version of the Fisher information about $t$ brought by $\mathbf{x}.$ From (%
\ref{5}), we get that the rate of increase of the mean fitness (which is the
variance in relative fitness) identifies with this Fisher information 
\begin{equation}
\frac{\Delta w\left( \mathbf{x}\right) }{w\left( \mathbf{x}\right) }=%
\overline{\sigma }^{2}\left( \mathbf{x}\right) =I_{\mathbf{x}}\left(
t\right) >0.\text{ }\blacklozenge  \label{5a}
\end{equation}

$\left( ii\right) $ When $w_{k}=1+sa_{k}$ as in the selection example, the
variance in relative fitness reads 
\begin{equation*}
\overline{\sigma }^{2}\left( \mathbf{x}\right) =\left( \frac{s}{1+sa\left( 
\mathbf{x}\right) }\right) ^{2}\sum_{k}x_{k}\left( a_{k}-a\left( \mathbf{x}%
\right) \right) ^{2}.\text{ }\blacklozenge
\end{equation*}

\subsection{Single locus: diploid population with $K$ alleles}

We now run into similar considerations but with diploid populations whose
genetical information governing their developments is carried by pairs of
chromosomes. When considering the estimation problem, to avoid overburden
notations that would blur the exposition, we shall limit ourselves to the
special one-parameter fitness model where a single selection parameter $s$
is unknown. Under this hypothesis, the estimation problem is over-simplified
because it avoids estimating the full fitness array that would lead to
additional notational and technical difficulties due to multidimensionality.

\textbf{Joint EG dynamics.} Let $w_{k,l}>0$, $k,l=1,...,K$ stand for the
absolute fitness of the genotypes $A_{k}A_{l}$ attached to a single locus$.$
Assume $w_{k,l}=w_{l,k}$. Let $W$ be the symmetric fitness matrix with $k,l-$%
entry $w_{k,l}$. Assume the current frequency distribution at time $t$ of
the genotypes $A_{k}A_{l}$ is given by $x_{k,l}.$ Let $X$ be the frequencies
array with $k,l-$entry $x_{k,l}$. The joint EG dynamics in the diploid case
is given by the updating: 
\begin{equation}
x_{k,l}^{\prime }=x_{k,l}\frac{w_{k,l}}{w\left( X\right) }  \label{6}
\end{equation}
where the mean fitness $w$ now is given by: $w\left( X\right)
=\sum_{k,l}x_{k,l}w_{k,l}.$ Define the relative fitness of the genotype $%
A_{k}A_{l}$ by: $\overline{w}_{k,l}\left( X\right) :=\frac{w_{k,l}}{w\left(
X\right) }$ and let $\overline{W}\left( X\right) $ be the matrix with
entries $\overline{w}_{k,l}\left( X\right) $. Then the joint EG dynamics
takes the matrix form: 
\begin{equation*}
X^{\prime }=X\circ \overline{W}\left( X\right) =\overline{W}\left( X\right)
\circ X
\end{equation*}
where $\circ $ stands for the (commutative) Hadamard product of matrices.

Let $J$ be the $K\times K$ flat matrix whose entries are all $1$. Then 
\begin{equation*}
\Delta X:=X^{\prime }-X=\left( X-J\right) \circ \overline{W}\left( X\right) =%
\overline{W}\left( X\right) \circ \left( X-J\right) .
\end{equation*}
We shall also let 
\begin{equation}
\sigma ^{2}\left( X\right) =\sum_{k,l=1}^{K}x_{k,l}\left( w_{k,l}-w\left(
X\right) \right) ^{2}  \label{7}
\end{equation}
stand for the genotypic variance in absolute fitness and 
\begin{equation}
\overline{\sigma }^{2}\left( X\right) =\sum_{k,l=1}^{K}x_{k,l}\left( 
\overline{w}_{k,l}\left( X\right) -1\right) ^{2}=\sigma ^{2}\left( X\right)
/w\left( X\right) ^{2}  \label{8}
\end{equation}
will stand for the diploid variance in relative fitness.

Consider the problem of evaluating the increase of the mean fitness. We have 
\begin{equation}
\Delta w\left( X\right) =\sum_{k,l}\Delta
x_{k,l}w_{k,l}=\sum_{k,l}x_{k,l}\left( \frac{w_{k,l}^{2}}{w\left( X\right) }%
-w_{k,l}\right) =w\left( X\right) \overline{\sigma }^{2}\left( X\right) >0
\label{9}
\end{equation}
with a relative rate of increase: $\Delta w\left( X\right) /w\left( X\right)
=\overline{\sigma }^{2}\left( X\right) .$ This is the full diploid version
of the FTNS.\newline

\textbf{Marginal allelic dynamics.} Assuming a Hardy-Weinberg equilibrium,
the frequency distribution at time $t,$ say $x_{k,l}$, of the genotypes $%
A_{k}A_{l}$ is given by: $x_{k,l}=x_{k}x_{l}$ where $x_{k}=\sum_{l}x_{k,l}$
is the marginal frequency of allele $A_{k}$ in the whole genotypic population%
$.$ The whole frequency information is now enclosed within $\mathbf{x}%
:=x_{k},$ $k=1,...,K.$ For instance, the mean fitness $w$ now is given by
the quadratic form: $w\left( \mathbf{x}\right) =\sum_{k,l}x_{k}x_{l}w_{k,l}=%
\mathbf{x}^{*}W\mathbf{x}$ with $\mathbf{x}^{*}$ the transposed line vector
of the column vector $\mathbf{x}=X\mathbf{1}$ ($\mathbf{1}$ the unit $K$%
-vector). We shall also let 
\begin{equation}
\sigma ^{2}\left( \mathbf{x}\right) =\sum_{k,l=1}^{K}x_{k}x_{l}\left(
w_{k,l}-w\left( \mathbf{x}\right) \right) ^{2}  \label{10}
\end{equation}
stand for the genotypic variance in absolute fitness and 
\begin{equation*}
\overline{\sigma }^{2}\left( \mathbf{x}\right)
=\sum_{k,l=1}^{K}x_{k}x_{l}\left( \overline{w}_{k,l}\left( \mathbf{x}\right)
-1\right) ^{2}=\sigma ^{2}\left( \mathbf{x}\right) /w\left( \mathbf{x}%
\right) ^{2}
\end{equation*}
will stand for the diploid variance in relative fitness with $\overline{w}%
_{k,l}\left( \mathbf{x}\right) :=w_{k,l}/w\left( \mathbf{x}\right) $ the
relative fitnesses. These quantities may now simply be indexed by $\mathbf{x}
$.

Before we come to the diploid marginal EG dynamics, let us make the
following remarks. Let 
\begin{equation}
S^{2}\left( \mathbf{\alpha }\right) :=\sum_{k,l=1}^{K}x_{k}x_{l}\left(
w_{k,l}-w\left( \mathbf{x}\right) -\alpha _{k}-\alpha _{l}\right) ^{2}.
\label{11}
\end{equation}
The values of $\mathbf{\alpha }$ minimizing $S^{2}\left( \mathbf{\alpha }%
\right) $ are easily seen to be $\mathbf{\alpha }^{*}=\alpha
_{k}^{*}=w_{k}\left( \mathbf{x}\right) -w\left( \mathbf{x}\right) ,$ $%
k=1,...,K.$ We shall let 
\begin{equation}
\sigma _{D}^{2}\left( \mathbf{x}\right) :=S^{2}\left( \mathbf{\alpha }%
^{*}\right) =\sum_{k,l=1}^{K}x_{k}x_{l}\left( w_{k,l}-\left( w_{k}\left( 
\mathbf{x}\right) +w_{l}\left( \mathbf{x}\right) -w\left( \mathbf{x}\right)
\right) \right) ^{2}  \label{12}
\end{equation}
and call it the dominance variance$.$ Then we get 
\begin{equation}
\sigma ^{2}\left( \mathbf{x}\right) =\sigma _{D}^{2}\left( \mathbf{x}\right)
+\sigma _{A}^{2}\left( \mathbf{x}\right) .  \label{13}
\end{equation}
The variance $\sigma _{A}^{2}\left( \mathbf{x}\right) $ is that component of
the total variance in absolute fitness of the genotypes which can be
explained additively by the alleles constituting those genotypes, \cite{C}.
Indeed, we can easily check that 
\begin{equation}
\sigma _{A}^{2}\left( \mathbf{x}\right) =2\sum_{k=1}^{K}x_{k}\left(
w_{k}\left( \mathbf{x}\right) -w\left( \mathbf{x}\right) \right) ^{2}.
\label{14}
\end{equation}
The number $\sigma _{A}^{2}\left( \mathbf{x}\right) /2$ can be interpreted
in terms of the fitness covariance between parent and offspring in the
updating step (see Ewens, \cite{Ew}, p. $7$).

The residual part $\sigma _{D}^{2}\left( \mathbf{x}\right) $ is that
component of $\sigma ^{2}\left( \mathbf{x}\right) $ which can be explained
by the interactions pertaining to dominance between the alleles forming the
genotypes.\newline

Consider now the update of the allelic marginal frequencies $\mathbf{x}$
themselves. If we first define the frequency-dependent marginal fitness of $%
A_{k}$ by $w_{k}\left( \mathbf{x}\right) =\left( W\mathbf{x}\right)
_{k}:=\sum_{l}w_{k,l}x_{l}$, the marginal dynamics is given as in (\ref{4})
by: 
\begin{equation}
x_{k}^{\prime }=x_{k}\overline{w}_{k}\left( \mathbf{x}\right) =:p_{k}\left( 
\mathbf{x}\right) \text{, }k=1,...,K  \label{15}
\end{equation}
where now: $\overline{w}_{k}\left( \mathbf{x}\right) :=\frac{w_{k}\left( 
\mathbf{x}\right) }{w\left( \mathbf{x}\right) }$ is the relative fitness of $%
A_{k}.$ In vector form 
\begin{equation*}
\mathbf{x}^{\prime }=\frac{D_{\mathbf{x}}W\mathbf{x}}{\mathbf{x}^{*}W\mathbf{%
x}}=D_{\overline{\mathbf{w}}\left( \mathbf{x}\right) }\mathbf{x,}
\end{equation*}
where $\overline{\mathbf{w}}\left( \mathbf{x}\right) :=\overline{w}%
_{k}\left( \mathbf{x}\right) ,$ $k=1,...,K$. Again, the mean fitness $%
w\left( \mathbf{x}\right) ,$ as a Lyapounov function, increases as time
passes by. We indeed have 
\begin{equation*}
\Delta w\left( \mathbf{x}\right) =w\left( \mathbf{x}^{\prime }\right)
-w\left( \mathbf{x}\right) =\sum_{k,l}x_{k}\overline{w}_{k}\left( \mathbf{x}%
\right) w_{k,l}x_{l}\overline{w}_{l}\left( \mathbf{x}\right)
-\sum_{k,l}x_{k}w_{k,l}x_{l}>0,
\end{equation*}
because, defining $0<X\left( \mathbf{x}\right) :=\sum_{k,l}x_{k}\left( 1-%
\overline{w}_{k}\left( \mathbf{x}\right) \right) w_{k,l}\left( 1-\overline{w}%
_{l}\left( \mathbf{x}\right) \right) x_{l}$, we have 
\begin{equation*}
\Delta w\left( \mathbf{x}\right) =X\left( \mathbf{x}\right) +\frac{2}{%
w\left( \mathbf{x}\right) }\left( \sum_{k}x_{k}w_{k}\left( \mathbf{x}\right)
^{2}-w\left( \mathbf{x}\right) ^{2}\right) >0.
\end{equation*}
Its partial rate of increase due to frequency shifts only is 
\begin{equation*}
\frac{\Delta _{P}w\left( \mathbf{x}\right) }{w\left( \mathbf{x}\right) }:=%
\frac{\sum_{k}\Delta x_{k}w_{k}\left( \mathbf{x}\right) }{w\left( \mathbf{x}%
\right) }.
\end{equation*}
This quantity is half the allelic variance in relative fitness $\sigma
_{A}^{2}\left( \mathbf{x}\right) /\left( 2w\left( \mathbf{x}\right)
^{2}\right) =\overline{\sigma }_{A}^{2}\left( \mathbf{x}\right) /2$. Indeed, 
\begin{equation}
\frac{\Delta _{P}w\left( \mathbf{x}\right) }{w\left( \mathbf{x}\right) }%
=\sum_{k}x_{k}\left( \overline{w}_{k}\left( \mathbf{x}\right) -1\right)
^{2}=\sum_{k}\frac{\left( \Delta x_{k}\right) ^{2}}{x_{k}}=\overline{\sigma }%
_{A}^{2}\left( \mathbf{x}\right) /2.  \label{16}
\end{equation}
The mean fitness increase phenomena (either global or partial) occur till
the EG dynamics reaches an equilibrium state. In the diploid case, this
dynamics can have more complex equilibrium points, satisfying $w_{k}\left( 
\mathbf{x}_{eq}\right) =w_{1}\left( \mathbf{x}_{eq}\right) $, $k=2,...,K$
and $\sum_{l}x_{eq,l}=1.$ In particular, a stable internal (polymorphic)
equilibrium state can exist, a necessary and sufficient condition being that 
$W$ has exactly one strictly positive dominant eigenvalue and at least one
strictly negative eigenvalue (see Kingman, \cite{K}) or else that the
sequence of principal minors of $W$ alternates in sign. An internal
polymorphic equilibrium state is asymptotically stable iff it is an isolated
local maximum of the mean fitness. If this is the case, there is a unique $%
\mathbf{z}>0$ for which $W\mathbf{z}=\mathbf{1}$ and the equilibrium
polymorphic state is $\mathbf{x}_{eq}=\mathbf{z/}\left| \mathbf{z}\right| .$
Moreover, starting from any initial condition in the interior of $S_{K},$
all trajectories are attracted by this $\mathbf{x}_{eq}$. When there is no
such unique globally stable polymorphic equilibrium, all trajectories will
still converge but perhaps to a local equilibrium state where some alleles
get extinct.

Except for the fact that the mean fitness now is a quadratic form in $%
\mathbf{x}$ and that the marginal fitness of $A_{k}$ now is
frequency-dependent, depending linearly on $\mathbf{x}$, as far as the
marginal frequencies are concerned, the updating formalism (\ref{15}) in the
diploid case looks very similar to the one in (\ref{4}) describing the
haploid case. In the diploid case, assuming fitnesses are multiplicative,
say with $W_{k,l}=w_{k}w_{l},$ then $\overline{w}_{k}\left( \mathbf{x}%
\right) :=\frac{w_{k}\left( \mathbf{x}\right) }{\mathbf{x}^{*}W\mathbf{x}}=%
\frac{w_{k}}{\sum_{l}w_{l}x_{l}}$ and the dynamics (\ref{15}) boils down to (%
\ref{4}). However, the mean fitness in this case is $w\left( \mathbf{x}%
\right) =\left( \sum_{l}w_{l}x_{l}\right) ^{2}$ and not $\sum_{l}w_{l}x_{l}$
as in the haploid case.\newline

\textbf{Example (selection).} In general the whole fitness matrix is
unknown. In some cases, only one selection parameter $s$ is to be determined
(estimated from data). Assume indeed $w_{k,l}=1+sa_{k,l}$ where $%
a_{k,l}=a_{l,k}\in \left[ 0,1\right] $ are known and $s>-1.$ A natural
choice could be $a_{k,l}=\frac{K-k}{K-1}\frac{K-l}{K-1}$, with $a_{1,1}=1$
and $a_{K,K}=0$. Or else: $a_{k,l}=\left( K/k-1\right) \left( K/l-1\right)
/\left( K-1\right) ^{2}$. The simple popular model $a_{k,l}=\delta _{k,l}$
is also of wide use in this context (see \cite{Ew}, p. $53$ or \cite{K1} p. $%
14$).

Then $W=J+sA$ where $A$ is a known matrix whose $k,l-$entry is $a_{k,l}$.
With $a_{k}\left( \mathbf{x}\right) =\sum_{l}a_{k,l}x_{l}$ and $a\left( 
\mathbf{x}\right) =\mathbf{x}^{*}A\mathbf{x,}$ the EG dynamics reads 
\begin{equation}
x_{k}^{\prime }=x_{k}\frac{1+sa_{k}\left( \mathbf{x}\right) }{1+sa\left( 
\mathbf{x}\right) }\text{, }k=1,...,K.  \label{17}
\end{equation}
The allelic variance in relative fitness reads 
\begin{equation}
\overline{\sigma }_{A}^{2}\left( \mathbf{x}\right) /2=\left( \frac{s}{%
1+sa\left( \mathbf{x}\right) }\right) ^{2}\sum_{k}x_{k}\left( a_{k}\left( 
\mathbf{x}\right) -a\left( \mathbf{x}\right) \right) ^{2}.\text{ }%
\blacklozenge  \label{18}
\end{equation}

\textbf{Remarks:}

$\left( i\right) $ There is an alternative vectorial representation of the
dynamics (\ref{15}) and (\ref{17}). Define the symmetric positive-definite
matrix $Q\left( \mathbf{x}\right) $ with quadratic entries in the
frequencies: 
\begin{equation*}
Q\left( \mathbf{x}\right) _{k,l}=x_{k}\left( \delta _{k,l}-x_{l}\right) .
\end{equation*}
Introduce the column vector of the relative fitnesses: $\overline{\mathbf{w}}%
\left( \mathbf{x}\right) =\overline{w}_{k}\left( \mathbf{x}\right) ,$ $%
k=1,...,K$ (with $\overline{\mathbf{w}}\left( \mathbf{x}\right) =:\nabla
V\left( \mathbf{x}\right) =\frac{1}{2}\nabla \log w\left( \mathbf{x}\right)
, $ half the gradient of the logarithm of mean fitness)$.$ Then, (\ref{15})
may be recast as the gradient-like replicator dynamics: 
\begin{equation}
\Delta \mathbf{x}=Q\left( \mathbf{x}\right) \overline{\mathbf{w}}\left( 
\mathbf{x}\right) =\frac{1}{w\left( \mathbf{x}\right) }Q\left( \mathbf{x}%
\right) W\mathbf{x}=Q\left( \mathbf{x}\right) \nabla V\left( \mathbf{x}%
\right) ,  \label{18c}
\end{equation}
with $\left| \Delta \mathbf{x}\right| =\mathbf{1}^{*}\Delta \mathbf{x}=0$ as
a result of $\mathbf{1}^{*}Q\left( \mathbf{x}\right) =\mathbf{0}^{*}$. Note 
\begin{equation*}
\nabla V\left( \mathbf{x}\right) ^{*}\Delta \mathbf{x}=\nabla V\left( 
\mathbf{x}\right) ^{*}Q\left( \mathbf{x}\right) \nabla V\left( \mathbf{x}%
\right) \geq 0.
\end{equation*}
In the selection case when $w_{k}\left( \mathbf{x}\right) =1+sa_{k}\left( 
\mathbf{x}\right) $, using $Q\left( \mathbf{x}\right) \mathbf{1}=\mathbf{0}:$%
\begin{equation*}
\Delta \mathbf{x}=Q\left( \mathbf{x}\right) \left( \overline{\mathbf{w}}%
\left( \mathbf{x}\right) -\mathbf{1}\right) =\frac{s}{1+sa\left( \mathbf{x}%
\right) }Q\left( \mathbf{x}\right) A\mathbf{x}.\text{ }\blacklozenge
\end{equation*}

$\left( ii\right) $ Although we shall not run into details pertaining to the
continuous-time setting, let us say a few words on this particular aspect.
In continuous-time $t\geq 0$, the dynamics of $x_{k}:=x_{k}\left( t\right) $
is 
\begin{equation*}
\overset{\cdot }{x}_{k}=x_{k}\left( w_{k}\left( \mathbf{x}\right) -w\left( 
\mathbf{x}\right) \right) \text{, }k=1,...,K
\end{equation*}
where the `dot' is the time-derivative. The growth rate is driven by the
average excess in mean fitness $w_{k}\left( \mathbf{x}\right) -w\left( 
\mathbf{x}\right) .$ Alternatively, the dynamics on $S_{K}$ is $\overset{%
\cdot }{\mathbf{x}}$ $=Q\left( \mathbf{x}\right) W\mathbf{x}$ with $\frac{d}{%
dt}\left| \mathbf{x}\right| =0$ because $\left| \mathbf{x}\right| =\mathbf{1}%
^{*}\mathbf{x}=\left\langle \mathbf{1,x}\right\rangle $ and $\mathbf{1}%
^{*}Q\left( \mathbf{x}\right) =Q\left( \mathbf{x}\right) \mathbf{1}=\mathbf{0%
}.$

In the special selection case, $\overset{\cdot }{\mathbf{x}}=sQ\left( 
\mathbf{x}\right) A\mathbf{x}$ or 
\begin{equation*}
\overset{\cdot }{x}_{k}=sx_{k}\left( a_{k}\left( \mathbf{x}\right) -a\left( 
\mathbf{x}\right) \right) \text{, }k=1,...,K.
\end{equation*}
In this case, the positive quantity 
\begin{equation*}
\sum_{k=1}^{K}\frac{\overset{}{\left( \overset{\cdot }{x}_{k}\right) ^{2}}}{%
x_{k}}=\sum_{k=1}^{K}x_{k}\overset{}{\left[ \frac{d\left( \log x_{k}\right) 
}{dt}\right] ^{2}}=s^{2}\sum_{k=1}^{K}x_{k}\left( a_{k}\left( \mathbf{x}%
\right) -a\left( \mathbf{x}\right) \right) ^{2}
\end{equation*}
may be viewed as the familiar Fisher information $I_{\mathbf{x}}:=I_{\mathbf{%
x}}\left( t\right) $ of the frequency distribution $\mathbf{x,}$ as a
discrete probability distribution parameterized by continuous time $t\mathbf{%
.}$ One can check that, if the mean fitness is $w\left( \mathbf{x}\right)
=1+sa\left( \mathbf{x}\right) $, then 
\begin{equation*}
\overset{\cdot }{w}\left( \mathbf{x}\right) =s\overset{\cdot }{a}\left( 
\mathbf{x}\right) =2I_{\mathbf{x}}.
\end{equation*}
So the time derivative of $w\left( \mathbf{x}\right) $ coincides with twice
this Fisher information. This constitutes the diploid continuous-time
version of (\ref{5a}).

Defining the dimensionless parameter $\theta :=st$ and looking at the
time-changed frequencies $\pi _{k}\left( \theta \right) :=x_{k}\left( \theta
/s\right) $, we get the $s-$free dynamics 
\begin{equation*}
\overset{\cdot }{\pi }_{k}=\pi _{k}\left( a_{k}\left( \mathbf{\pi }\right)
-a\left( \mathbf{\pi }\right) \right) \text{, }k=1,...,K,
\end{equation*}
where the `dot' now is the derivative with respect to $\theta $. Clearly, $%
\overset{\cdot }{w}\left( \mathbf{\pi }\right) =$ $\overset{\cdot }{a}\left( 
\mathbf{\pi }\right) =2I_{\mathbf{\pi }}\left( \theta \right) >0.$ $%
\blacklozenge $\newline

\textbf{Partial change of mean fitness. }Let us return to the joint EG
dynamics where no hypothesis on mating was made and consider the full mean
fitness 
\begin{equation}
w\left( X\right) =\sum_{k,l}x_{k,l}w_{k,l}.  \label{18a}
\end{equation}
Define $\alpha _{k}=w_{k}\left( X\right) -w\left( X\right) $ where $%
w_{k}\left( X\right) :=\sum_{l}x_{k,l}w_{k,l}/x_{k}$ and $%
x_{k}:=\sum_{l}x_{k,l}$ is the marginal frequency of $A_{k}.$ Replace the
expression (\ref{18a}) by the equally correct 
\begin{equation}
w\left( X\right) =\sum_{k,l}x_{k,l}\left( w\left( X\right) +\alpha
_{k}+\alpha _{l}\right) ,  \label{18b}
\end{equation}
suggesting that the fitnesses of the genotypes $A_{k}A_{l}$ would rather be $%
w_{k,l}^{\left( \alpha \right) }:=w\left( X\right) +\alpha _{k}+\alpha _{l}$.

Define the partial change, say $\Delta _{P}$, of mean fitness as 
\begin{equation*}
\Delta _{P}w\left( X\right) :=\sum_{k,l}\Delta x_{k,l}\left( w\left(
X\right) +\alpha _{k}+\alpha _{l}\right)
\end{equation*}
where only a variation in the frequency term is considered. After some
elementary algebra, we get 
\begin{equation*}
\Delta _{P}w\left( X\right) =\sum_{k,l}\Delta x_{k,l}\left( \alpha
_{k}+\alpha _{l}\right) =2\sum_{k}\alpha _{k}\sum_{l}\Delta x_{k,l}
\end{equation*}
\begin{equation*}
=2\sum_{k}\alpha _{k}\Delta x_{k}=:\sigma _{A}^{2}\left( X\right) /w\left(
X\right) ,
\end{equation*}
leading to a partial rate of increase $\Delta _{P}w\left( X\right) /w\left(
X\right) =\overline{\sigma }_{A}^{2}\left( X\right) $ which is similar to (%
\ref{16}). In the alternative Castilloux-Lessard interpretation of this
phenomenon, \cite{CL}, defining $\Delta x_{k,l}^{\left( \alpha \right)
}:=\Delta x_{k,l}\frac{\alpha _{k}+\alpha _{l}}{w\left( X\right) }$ and
observing 
\begin{equation*}
\Delta _{P}w\left( X\right) =\sum_{k,l}\Delta x_{k,l}^{\left( \alpha \right)
}w_{k,l},
\end{equation*}
the partial change involves an allele-based modification of the genotype
frequencies while the genotype fitnesses $w_{k,l}$ are kept unchanged.

\subsection{Estimation of $s$}

We now switch to the announced estimation of $s$ problem which seems to be
new.\newline

Assume the current and updated frequencies $\mathbf{x}$ and $\mathbf{x}%
^{\prime }$ are being observed at some times $t,$ $t+1$. We wish to use this
information to estimate the unknown value of $s.$ Let first $s_{k}^{*}$ be
the estimate of $s$ which explains the observable $\left( \mathbf{x;}%
x_{k}^{\prime }\right) $ at best$.$ Clearly, from (\ref{17}), 
\begin{equation}
s_{k}^{*}=\frac{x_{k}^{\prime }-x_{k}}{a_{k}\left( \mathbf{x}\right)
x_{k}-x_{k}^{\prime }a\left( \mathbf{x}\right) }  \label{19}
\end{equation}
does the job. A natural estimator $s^{*}=s^{*}\left( \mathbf{x;x}^{\prime
}\right) $ of $s$ which explains best the observable $\left( \mathbf{x;x}%
^{\prime }\right) $ is: 
\begin{equation*}
s^{*}=\arg \min_{s}\sum_{k}x_{k}x_{k}^{\prime }\left( s-s_{k}^{*}\right) ^{2}
\end{equation*}
which is: 
\begin{equation}
s^{*}=\frac{1}{\sum_{l}x_{l}x_{l}^{\prime }}\sum_{k}x_{k}x_{k}^{\prime
}s_{k}^{*},  \label{20}
\end{equation}
a weighted average of the $s_{k}^{*}$ attributing more credit to $s_{k}^{*}$
when $x_{k}x_{k}^{\prime }$ is largest$.$\newline

Sometimes, the $x_{k},$ $x_{k}^{\prime }$ are not directly observed. Rather,
what is observed is the scalar output: 
\begin{equation*}
y:=h\left( \mathbf{x}\right) :=\sum_{k}x_{k}h_{k}\left( \mathbf{x}\right)
\end{equation*}
for some known family of measurements $h_{k}\left( \mathbf{x}\right) $, $%
k=1,...,K$ given the process $\mathbf{x}$ is in state $k.$ Simple but
important examples are $h_{k}\left( \mathbf{x}\right) =x_{k}^{\alpha -1}$ ($%
\alpha >1$) in which case, $y=h\left( \mathbf{x}\right)
=\sum_{k}x_{k}^{\alpha }$ is $\alpha -$homozygosity (typically $\alpha =2$),
or $h_{k}\left( \mathbf{x}\right) =-\log x_{k}$ in which case, $y=h\left( 
\mathbf{x}\right) =-\sum_{k}x_{k}\log x_{k}$ is the Shannon entropy of the
frequency distribution.

Let $\kappa $ be a discrete random variable with $\mathbf{P}\left( \kappa
=k\right) =x_{k},$ $k=1,...,K$ so that $y:=\mathbf{E}\left( h_{\kappa
}\left( \mathbf{x}\right) \right) ,$ the mathematical expectation with
respect to $\kappa $s law$.$ From (\ref{17}), we have: 
\begin{equation*}
y^{\prime }=h\left( \mathbf{x}^{\prime }\right) :=\sum_{k}x_{k}^{\prime
}h_{k}\left( \mathbf{x}^{\prime }\right) =\sum_{k}\Delta x_{k}h_{k}\left( 
\mathbf{x}^{\prime }\right) +\sum_{k}x_{k}h_{k}\left( \mathbf{x}^{\prime
}\right)
\end{equation*}
\begin{equation*}
=\frac{s}{1+sa\left( \mathbf{x}\right) }\sum_{k}x_{k}h_{k}\left( \mathbf{x}%
^{\prime }\right) \left( a_{k}\left( \mathbf{x}\right) -a\left( \mathbf{x}%
\right) \right) +\mathbf{E}\left( h_{\kappa }\left( \mathbf{x}^{\prime
}\right) \right)
\end{equation*}
\begin{equation*}
=\frac{s}{1+sa\left( \mathbf{x}\right) }\text{\textbf{Cov}}\left( h_{\kappa
}\left( \mathbf{x}^{\prime }\right) ,a_{\kappa }\left( \mathbf{x}\right)
\right) +\mathbf{E}\left( h_{\kappa }\left( \mathbf{x}^{\prime }\right)
\right) .
\end{equation*}
Therefore, the observed shift in the measurement is 
\begin{equation}
\Delta y=\frac{s}{1+sa\left( \mathbf{x}\right) }\text{\textbf{Cov}}\left(
h_{\kappa }\left( \mathbf{x}^{\prime }\right) ,a_{\kappa }\left( \mathbf{x}%
\right) \right) +\mathbf{E}\left( \Delta h_{\kappa }\left( \mathbf{x}\right)
\right)  \label{20a}
\end{equation}
and an estimate $s^{*}$ based on $\Delta y$ can immediately be written down
by mere solving the above equation (\ref{20a}) which is reminiscent of a
Price equation (see \cite{F1}). It involves two terms, one which is related
to the correlation between the measurement at $t+1$ and the fitness function
at $t$ that has to do with the frequency shift only, another related to the
induced change of the character value only.

\section{EG theory: the stochastic point of view}

With $\overline{w}_{k}\left( \mathbf{x}\right) :=\frac{w_{k}\left( \mathbf{x}%
\right) }{w\left( \mathbf{x}\right) },$ let $p_{k}\left( \mathbf{x}\right)
:=x_{k}\overline{w}_{k}\left( \mathbf{x}\right) $, $k=1,...,K$ with $%
\sum_{k}p_{k}\left( \mathbf{x}\right) =1$ be defined as in the previous
Section either from allelic or genotypic fitnesses. We shall assume
throughout that the special selection model assumptions: $w_{k}\left( 
\mathbf{x}\right) =1+sa_{k}\left( \mathbf{x}\right) $ and $w\left( \mathbf{x}%
\right) =1+sa\left( \mathbf{x}\right) $ are at stake$.$

\subsection{The Wright-Fisher model}

We start considering similar problems under the Wright-Fisher model.\newline

\textbf{The Model and first properties.} Consider an allelic population with
constant size $N.$ In the haploid (diploid) case, $N$ is (twice) the number
of real individuals$.$ Let $\mathbf{i}:=i_{k}$ and $\mathbf{i}^{\prime
}:=i_{k}^{\prime },$ $k=1,...,K$ be two vectors of integers quantifying the
size of the allelic populations at two consecutive generations $t$ and $t+1$%
. With $\left| \mathbf{i}\right| =\sum_{k}i_{k}$, therefore $\left| \frac{%
\mathbf{i}}{N}\right| =\left| \frac{\mathbf{i}^{\prime }}{N}\right| =1$ on $%
S_{K}$ $.$ Suppose the stochastic EG dynamics now is given by a Markov chain
whose one-step transition matrix $P$ from states $\mathbf{I}=\mathbf{i}$ to $%
\mathbf{I}^{\prime }=\mathbf{i}^{\prime }$ is given by the multinomial
Wright-Fisher (WF) model 
\begin{equation}
\Bbb{P}\left( \mathbf{I}_{t+1}^{\prime }=\mathbf{i}^{\prime }\mid \mathbf{I}%
_{t}=\mathbf{i}\right) =:P\left( \mathbf{i},\mathbf{i}^{\prime }\right) =%
\binom{N}{i_{1}^{\prime }\cdots i_{K}^{\prime }}\prod_{k=1}^{K}p_{k}\left( 
\frac{\mathbf{i}}{N}\right) ^{i_{k}^{\prime }}.  \label{21}
\end{equation}
The state-space dimension of this Markov chain is $\binom{N+1}{K-1}$ (the
number of compositions of integer $N$ into $K$ non-negative parts).

Let $\mathbf{e}_{l}$ be the $K-$null vector except for its $l-$th entry
which is $1.$ The extremal states $S_{K}^{*}:=\left\{ \mathbf{i}_{l}^{*}:=N%
\mathbf{e}_{l},l=1,...,K\right\} ,$ are all absorbing for this Markov chain
because $p_{k}\left( \frac{\mathbf{i}_{l}^{*}}{N}\right) =\delta _{k,l}$.
Under our assumptions, the chain is not recurrent. Depending on the initial
condition, say $\mathbf{i}_{0}$, the chain will necessarily end up in one of
the extremal states $\mathbf{i}_{l}^{*}$, with some probability, say $\pi
_{l}\left( \mathbf{i}_{0}\right) $, which can be computed as follows. Let $%
\mathbf{\pi }_{l}:=\pi _{l}\left( \mathbf{i}\right) $, $\mathbf{i}\in S_{K}$
be an harmonic function of the WF Markov chain, solution to: 
\begin{equation}
\left( P-I\right) \mathbf{\pi }_{l}=\mathbf{0}\text{ if }\mathbf{i}\in
S_{K}\backslash S_{K}^{*}\text{ and }\mathbf{\pi }_{l}=1\left( \mathbf{i}=N%
\mathbf{e}_{l}\right) \text{ if }\mathbf{i}\in S_{K}^{*}.  \label{21a}
\end{equation}
It satisfies 
\begin{equation*}
\Bbb{P}\left( \mathbf{I}_{\tau }=N\mathbf{e}_{l}\mid \mathbf{I}_{0}=\mathbf{i%
}_{0}\right) =\pi _{l}\left( \mathbf{i}_{0}\right) ,
\end{equation*}
where $\tau $ ($<\infty $ almost surely) is the random hitting time of $%
S_{K}^{*}$ for $\mathbf{I}_{t}$ and the $\pi _{l}\left( \mathbf{i}%
_{0}\right) $s are normalized so as $\sum_{l}\pi _{l}\left( \mathbf{i}%
_{0}\right) =1$. Thus $\pi _{l}\left( \mathbf{i}_{0}\right) $ is the
searched probability to end up in state $N\mathbf{e}_{l}$ starting from
state $\mathbf{i}_{0}$. In the same vein, the expected hitting time $\alpha
\left( \mathbf{i}_{0}\right) :=\Bbb{E}_{\mathbf{i}_{0}}\left( \tau \right) $
solves: 
\begin{eqnarray*}
\left( P-I\right) \mathbf{\alpha } &=&\mathbf{1}\text{, }\mathbf{i}/N\in
S_{K}\backslash S_{K}^{*} \\
\mathbf{\alpha } &=&0\text{, }\mathbf{i}/N\in S_{K}^{*}
\end{eqnarray*}
where $\mathbf{\alpha }:=\alpha \left( \mathbf{i}\right) $, $\mathbf{i}\in
S_{K}$. With $\mathbf{\pi }_{l}$ the solution to the above Dirichlet
problem, the equilibrium measure of the chain therefore is: 
\begin{equation*}
\pi _{eq}:=\sum_{l=1}^{K}\pi _{l}\left( \mathbf{i}_{0}\right) \delta _{%
\mathbf{i}_{l}^{*}},
\end{equation*}
which depends on $\mathbf{i}_{0}$. Unless some prior information on $\mathbf{%
i}_{0}$ is given, we may assume that $\mathbf{i}_{0}=\frac{N}{K}\mathbf{1}$
in which case one expects $\pi _{l}\left( \mathbf{i}_{0}\right) =1/K$ and $%
\pi _{eq}$ is uniform on the extremal states$.$

Necessarily, one allele will fixate and there is no polymorphic equilibrium
state even when dealing with diploid populations. Which allele and with what
probability will depend on the initial condition. Thanks to fluctuations,
the picture therefore looks very different from the one pertaining to the
deterministic theory. For analogies of this construction with statistical
physics, see \cite{SH}, \cite{S}.

The marginal transition matrix from $\mathbf{i}$ to $I_{k}^{\prime
}=i_{k}^{\prime }$ is binomial bin$\left( N,p_{k}\left( \frac{\mathbf{i}}{N}%
\right) \right) $: 
\begin{equation*}
P\left( \mathbf{i},i_{k}^{\prime }\right) =\binom{N}{i_{k}^{\prime }}%
p_{k}\left( \frac{\mathbf{i}}{N}\right) ^{{}}\left( 1-p_{k}\left( \frac{%
\mathbf{i}}{N}\right) \right) ^{N-i_{k}^{\prime }}.
\end{equation*}
With $p_{k}\left( \frac{\mathbf{i}}{N}\right) :=\frac{i_{k}}{N}\overline{w}%
_{k}\left( \frac{\mathbf{i}}{N}\right) $, given $\mathbf{I}=\mathbf{i}$, the 
$k-$th component $I_{k}^{\prime }$ of the updated state is now random with: 
\begin{equation*}
\Bbb{E}_{\mathbf{i}}\left( I_{k}^{\prime }\right) =Np_{k}\left( \frac{%
\mathbf{i}}{N}\right) \text{ and }\sigma _{\mathbf{i}}^{2}\left(
I_{k}^{\prime }\right) =Np_{k}\left( \frac{\mathbf{i}}{N}\right) \left(
1-p_{k}\left( \frac{\mathbf{i}}{N}\right) \right) .
\end{equation*}
\newline

\textbf{Mean fitness.} We shall introduce the random increment in absolute
mean fitness as 
\begin{equation}
\Delta w_{\mathbf{I}^{\prime }}\left( \frac{\mathbf{i}}{N}\right)
:=\sum_{k=1}^{K}\left( \frac{I_{k}^{\prime }}{N}-\frac{i_{k}}{N}\right)
w_{k}\left( \frac{\mathbf{i}}{N}\right) =s\sum_{k=1}^{K}\left( \frac{%
I_{k}^{\prime }}{N}-\frac{i_{k}}{N}\right) a_{k}\left( \frac{\mathbf{i}}{N}%
\right) .  \label{22}
\end{equation}
Dropping for notational ease the argument $\frac{\mathbf{i}}{N}$ appearing
in $\Delta w_{\mathbf{I}^{\prime }},$ $a_{k},$ $a$ and $p_{k}$, we get 
\begin{equation*}
\Bbb{E}_{\mathbf{i}}\Delta w_{\mathbf{I}^{\prime }}=s\sum_{k=1}^{K}\frac{%
i_{k}}{N}\left( \frac{1+sa_{k}}{1+sa}-1\right) a_{k}=\frac{s^{2}}{1+sa}%
\sum_{k=1}^{K}\frac{i_{k}}{N}\left( a_{k}-a\right) a_{k}
\end{equation*}
\begin{equation}
=\frac{s^{2}}{1+sa}\left[ \sum_{k=1}^{K}\frac{i_{k}}{N}a_{k}^{2}-a^{2}%
\right] >0.  \label{22a}
\end{equation}
The mean of the increment of the random absolute mean fitness is positive (a
random version of the FTNS). Its rate of increase is 
\begin{equation}
\frac{\Bbb{E}_{\mathbf{i}}\Delta w_{\mathbf{I}^{\prime }}\left( \frac{%
\mathbf{i}}{N}\right) }{w\left( \frac{\mathbf{i}}{N}\right) }=\left( \frac{s%
}{1+sa}\right) ^{2}\left[ \sum_{k=1}^{K}\frac{i_{k}}{N}a_{k}^{2}-a^{2}%
\right] ,  \label{23}
\end{equation}
involving the variance of the $a_{k}$s under the current frequency
distribution $\frac{i_{k}}{N},$ $k=1,...,K.$

Let us now compute the variance of $\Delta w_{\mathbf{I}^{\prime }}\left( 
\frac{\mathbf{i}}{N}\right) $. We get: 
\begin{equation}
\sigma _{\mathbf{i}}^{2}\left( \Delta w_{\mathbf{I}^{\prime }}\right)
=s^{2}\sigma _{\mathbf{i}}^{2}\left( \sum_{k=1}^{K}\frac{I_{k}^{\prime }}{N}%
a_{k}-a\right) =s^{2}\sigma _{\mathbf{i}}^{2}\left( \sum_{k=1}^{K}\frac{%
I_{k}^{\prime }}{N}a_{k}\right)  \label{24}
\end{equation}
\begin{equation*}
=s^{2}\left[ \Bbb{E}_{\mathbf{i}}\left( \sum_{k,k^{\prime }=1}^{K}\frac{%
I_{k}^{\prime }I_{k^{\prime }}^{\prime }}{N^{2}}a_{k}a_{k^{\prime }}\right)
-\left( \Bbb{E}_{\mathbf{i}}\left( \sum_{k=1}^{K}\frac{I_{k}^{\prime }}{N}%
a_{k}\right) \right) ^{2}\right] .
\end{equation*}
It is proportional to the variance of the weighted outcomes $\sum_{k=1}^{K}%
\frac{I_{k}^{\prime }}{N}a_{k}$ given $\mathbf{i.}$

Using $\Bbb{E}_{\mathbf{i}}\left( I_{k}^{\prime }I_{k^{\prime }}^{\prime
}\right) =N\left( N-1\right) p_{k}p_{k^{\prime }}$ and $\Bbb{E}_{\mathbf{i}%
}\left( I_{k}^{\prime 2}\right) =Np_{k}+N\left( N-1\right) p_{k}^{2},$ we
get 
\begin{equation*}
\sigma _{\mathbf{i}}^{2}\left( \Delta w_{\mathbf{I}^{\prime }}\right)
=s^{2}\left[ \sum_{k,k^{\prime }=1}^{K}\frac{N\left( N-1\right)
p_{k}p_{k^{\prime }}}{N^{2}}a_{k}a_{k^{\prime }}+N\sum_{k=1}^{K}\frac{p_{k}}{%
N^{2}}a_{k}^{2}-\left( \sum_{k=1}^{K}p_{k}a_{k}\right) ^{2}\right]
\end{equation*}
and so, 
\begin{equation}
\sigma _{\mathbf{i}}^{2}\left( \Delta w_{\mathbf{I}^{\prime }}\left( \frac{%
\mathbf{i}}{N}\right) \right) =\frac{s^{2}}{N}\left[
\sum_{k=1}^{K}p_{k}a_{k}^{2}-\left( \sum_{k=1}^{K}p_{k}a_{k}\right)
^{2}\right] ,  \label{25}
\end{equation}
again involving the variance of the $a_{k}$s but now under the updated mean
frequency distribution $p_{k}=\Bbb{E}_{i}\left( \frac{I_{k}^{\prime }}{N}%
\right) =\frac{i_{k}}{N}\frac{1+sa_{k}}{1+sa},$ $k=1,...,K.$

We conclude that 
\begin{equation}
\frac{\Bbb{E}_{\mathbf{i}}\Delta w_{\mathbf{I}^{\prime }}\left( \frac{%
\mathbf{i}}{N}\right) }{w\left( \frac{\mathbf{i}}{N}\right) }\nsim \sigma _{%
\mathbf{i}}^{2}\left( \Delta w_{\mathbf{I}^{\prime }}\left( \frac{\mathbf{i}%
}{N}\right) \right)  \label{26}
\end{equation}
as one might have expected from the analogies with the deterministic theory.

In fact, the full law of $\Delta w_{\mathbf{I}^{\prime }}\left( \frac{%
\mathbf{i}}{N}\right) $ can be computed and the large $N$ population limit
is worth investigating. Indeed, its Laplace-Stieltjes transform (LST) reads 
\begin{equation*}
\Bbb{E}_{\mathbf{i}}\left( e^{-\lambda \Delta w_{\mathbf{I}^{\prime }}\left( 
\frac{\mathbf{i}}{N}\right) }\right) =e^{\lambda sa}\Bbb{E}_{\mathbf{i}%
}\left( e^{-\frac{\lambda s}{N}\sum_{k=1}^{K}I_{k}^{\prime }a_{k}}\right)
=\left( \sum_{k=1}^{K}p_{k}e^{-\frac{\lambda s}{N}\left( a_{k}-a\right)
}\right) ^{N},
\end{equation*}
suggesting from large deviation theory that, if $i_{k}:=\left\lfloor
Nx_{k}\right\rfloor $, $k=1,...,K$%
\begin{equation*}
\Delta w_{\mathbf{I}^{\prime }}\left( \frac{\left\lfloor N\mathbf{x}%
\right\rfloor }{N}\right) \overset{a.s.}{\underset{N\uparrow \infty }{%
\rightarrow }}\Delta w_{\mathbf{I}^{\prime }}\left( \mathbf{x}\right)
=s\sum_{k=1}^{K}p_{k}\left( \mathbf{x}\right) \left( a_{k}\left( \mathbf{x}%
\right) -a\left( \mathbf{x}\right) \right)
\end{equation*}
\begin{equation*}
=\frac{s^{2}}{1+sa\left( \mathbf{x}\right) }\left(
\sum_{k=1}^{K}x_{k}a_{k}\left( \mathbf{x}\right) ^{2}-a\left( \mathbf{x}%
\right) ^{2}\right) ,
\end{equation*}
which is the deterministic value $\sigma _{A}^{2}\left( \mathbf{x}\right)
/\left( 2\left( 1+sa\left( \mathbf{x}\right) \right) \right) $ of the
marginal deterministic theory (\ref{14}).\newline

\textbf{Statistics}. We now suppose the WF Markov chain is in state $\mathbf{%
i}$, with $\mathbf{i\neq i}_{l}^{*}$ so that it has not yet reached any of
its equilibrium states. Based on the observation $\mathbf{i}$, we would like
to design estimators of the selection parameter $s$.

The log-likelihood of the model (\ref{21}) is 
\begin{equation*}
\log P\left( \mathbf{i},\mathbf{i}^{\prime }\right) =\log \binom{N}{%
i_{1}^{\prime }...i_{K}^{\prime }}+\sum_{k=1}^{K}i_{k}^{\prime }\log \left[ 
\frac{i_{k}}{N}w_{k}\left( \frac{\mathbf{i}}{N}\right) /w\left( \frac{%
\mathbf{i}}{N}\right) \right] .
\end{equation*}
If $w_{k}=1+sa_{k}$ and $w=1+sa$, its derivative with respect to $s$ is 
\begin{equation*}
\partial _{s}\log P\left( \mathbf{i},\mathbf{i}^{\prime }\right) =\partial
_{s}\left( \sum_{k=1}^{K}i_{k}^{\prime }\log \left( 1+sa_{k}\right) -N\log
\left( 1+sa\right) \right)
\end{equation*}
\begin{equation*}
=\sum_{k=1}^{K}i_{k}^{\prime }\frac{a_{k}}{1+sa_{k}}-N\frac{a}{1+sa}.
\end{equation*}
The value $s^{MLE}=s^{MLE}\left( \frac{\mathbf{i}}{N}\mathbf{,}\frac{\mathbf{%
i}^{\prime }}{N}\right) $ for which $\partial _{s}\log P\left( \mathbf{i},%
\mathbf{i}^{\prime }\right) =0$ is the Maximum Likelihood Estimator of $s$
given the observable $\left( \mathbf{i;i}^{\prime }\right) .$ It is given by
the implicit equation 
\begin{equation}
\sum_{k=1}^{K}\frac{i_{k}^{\prime }}{N}\frac{a_{k}\left( \frac{\mathbf{i}}{N}%
\right) }{1+s^{MLE}a_{k}\left( \frac{\mathbf{i}}{N}\right) }=\frac{a\left( 
\frac{\mathbf{i}}{N}\right) }{1+s^{MLE}a\left( \frac{\mathbf{i}}{N}\right) }.
\label{27}
\end{equation}
It is probably biased. Let us compute the Fisher information on $s$ enclosed
in the observation $\left( \mathbf{i};\mathbf{I}^{\prime }\right) $ which is 
\begin{equation}
I_{\mathbf{i}}\left( s\right) =\Bbb{E}_{\mathbf{i}}\left[ \left( \partial
_{s}\log P\left( \mathbf{i},\mathbf{I}^{\prime }\right) \right) ^{2}\right] .
\label{28}
\end{equation}
We get 
\begin{equation*}
I_{\mathbf{i}}\left( s\right) =\sum_{\mathbf{i}^{\prime }}P\left( \mathbf{i},%
\mathbf{i}^{\prime }\right) \left[ \sum_{k=1}^{K}i_{k}^{\prime }\frac{a_{k}}{%
1+sa_{k}}-\frac{Na}{1+sa}\right] ^{2}=\sigma _{\mathbf{i}}^{2}\left(
\sum_{k=1}^{K}I_{k}^{\prime }\frac{a_{k}}{1+sa_{k}}\right)
\end{equation*}
\begin{equation*}
=\sum_{k,k^{\prime }=1}^{K}\frac{a_{k}}{1+sa_{k}}\frac{a_{k^{\prime }}}{%
1+sa_{k^{\prime }}}\Bbb{E}_{\mathbf{i}}\left( I_{k}^{\prime }I_{k^{\prime
}}^{\prime }\right) -\left( \frac{Na}{1+sa}\right) ^{2}
\end{equation*}
and therefore 
\begin{equation}
I_{\mathbf{i}}\left( s\right) =\frac{N}{1+sa}\left[ \sum_{k=1}^{K}\frac{i_{k}%
}{N}\left( \frac{a_{k}^{2}}{1+sa_{k}}-\frac{a^{2}}{1+sa}\right) \right] .
\label{29}
\end{equation}
The Fisher information is exactly the variance of the weighted outcomes $%
\sum_{k=1}^{K}\frac{I_{k}^{\prime }}{N}\frac{a_{k}}{1+sa_{k}}$ given $%
\mathbf{i.}$ We conclude that 
\begin{equation}
\frac{\Bbb{E}_{\mathbf{i}}\Delta w_{\mathbf{I}^{\prime }}\left( \frac{%
\mathbf{i}}{N}\right) }{w}\nsim I_{\mathbf{i}}\left( s\right) .  \label{30}
\end{equation}
From the expression of the mean $\Bbb{E}_{\mathbf{i}}\left( I_{k}^{\prime
}\right) $%
\begin{equation*}
\Bbb{E}_{\mathbf{i}}\left( \frac{I_{k}^{\prime }}{N}\right) =\frac{i_{k}}{N}%
\frac{1+sa_{k}}{1+sa}
\end{equation*}
and, with $\left\langle i_{k}^{\prime }\right\rangle _{\mathbf{i}}=\frac{1}{n%
}\sum_{m=1}^{n}i_{k,m}^{\prime }$ the empirical average given $\mathbf{i}$
of $I_{k}^{\prime }$ based on a size-$n$ sample observation of $%
I_{k}^{\prime }$%
\begin{equation*}
s_{k}^{*}=\frac{\left\langle i_{k}^{\prime }\right\rangle _{\mathbf{i}}-i_{k}%
}{a_{k}\left( \frac{\mathbf{i}}{N}\right) i_{k}-\left\langle i_{k}^{\prime
}\right\rangle _{\mathbf{i}}a\left( \frac{\mathbf{i}}{N}\right) }
\end{equation*}
is a first moment estimator of $s$ explaining best $\left( \mathbf{i;}%
i_{k,m}^{\prime },m=1,...,n\right) $ and 
\begin{equation}
s^{*}=\frac{1}{\sum_{l}i_{l}\left\langle i_{l}^{\prime }\right\rangle _{%
\mathbf{i}}}\sum_{k}i_{k}\left\langle i_{k}^{\prime }\right\rangle _{\mathbf{%
i}}s_{k}^{*},  \label{31}
\end{equation}
a moment estimator of $s$ explaining best $\left( \mathbf{i;i}_{m}^{\prime
},m=1,...,n\right) $ whenever we are in possession of $n$ observed copies $%
\mathbf{i}^{\prime }$ of $\mathbf{I}^{\prime }$ based on the same $\mathbf{i}
$. This estimator is also biased.

With $\widehat{s}=s^{*}$ or $s^{MLE}$, let 
\begin{equation}
R_{\mathbf{i}}\left( \widehat{s},s\right) =\Bbb{E}_{\mathbf{i}}\left[ \left( 
\widehat{s}-s\right) ^{2}\right] =\sigma _{\mathbf{i}}^{2}\left( \widehat{s}%
\right) +\left( \Bbb{E}_{\mathbf{i}}\left( \widehat{s}\right) -s\right) ^{2}
\label{32}
\end{equation}
be the quadratic risk function associated with the estimator $\widehat{s}.$
By the Fr\'{e}chet-Darmois-Cramer Rao theorem, we have 
\begin{equation}
R_{\mathbf{i}}\left( \widehat{s},s\right) \geq \frac{1}{I_{\mathbf{i}}\left(
s\right) },  \label{33}
\end{equation}
where, in this classical interpretation of the Fisher information, $I_{%
\mathbf{i}}\left( s\right) ^{-1}$ appears as a universal lower bound of the
estimator quadratic error.\newline

Finally, we would like to stress that these considerations are also relevant
in the context of another fundamental stochastic model arising in the
context of evolutionary genetics. We shall give some elements of how to
proceed with this model presenting very different properties.

\subsection{The $K-$allele Moran model\textbf{\ }}

We now focus on the estimation problem under the Moran model.\newline

\textbf{The Model.} Let\textbf{\ }$\alpha ,\beta \in \left\{ 1,...,K\right\}
.$ In the Moran version of the stochastic evolution,\textbf{\ }given $%
\mathbf{I}_{t}=\mathbf{I}=\mathbf{i}$, the only accessible values of $%
\mathbf{I}^{\prime }$ are the neighboring states: $\mathbf{i}_{\alpha ,\beta
}^{\prime }:=\mathbf{i}+\mathbf{d}_{\alpha ,\beta }$ where $\mathbf{d}%
_{\alpha ,\beta }=\left( 0,..,0,-1,0,...,1,0,...,0\right) .$ Here $-1$ is in
position $\alpha $ and $1$ in position $\beta \neq \alpha $ corresponding to
the transfer of an individual from cell $\alpha $ to cell $\beta .$ The
Moran stochastic EG dynamics now is given by a Markov chain whose one-step
transition matrix $P$ from states $\mathbf{I}=\mathbf{i}$ to $\mathbf{I}%
^{\prime }=\mathbf{i}^{\prime }$ is: 
\begin{equation}
\Bbb{P}\left( \mathbf{I}_{t+1}=\mathbf{i}^{\prime }\mid \mathbf{I}_{t}=%
\mathbf{i}\right) =0\text{ if }\mathbf{i}^{\prime }\neq \mathbf{i}_{\alpha
,\beta }^{\prime }\text{ and}  \label{45}
\end{equation}
\begin{equation*}
\Bbb{P}\left( \mathbf{I}_{t+1}=\mathbf{i}_{\alpha ,\beta }^{\prime }\mid 
\mathbf{I}_{t}=\mathbf{i}\right) =:P\left( \mathbf{i},\mathbf{i}_{\alpha
,\beta }^{\prime }\right) =\frac{i_{\alpha }}{N}p_{\beta }\left( \frac{%
\mathbf{i}}{N}\right) ,
\end{equation*}
where $p_{\beta }\left( \frac{\mathbf{i}}{N}\right) $ is given by $p_{\beta
}\left( \frac{\mathbf{i}}{N}\right) :=\frac{i_{\beta }}{N}\overline{w}%
_{\beta }\left( \frac{\mathbf{i}}{N}\right) $.

Summing $P\left( \mathbf{i},\mathbf{i}_{\alpha ,\beta }^{\prime }\right) $
over $\alpha ,\beta $, $\beta \neq \alpha $ in (\ref{45}), we get the
holding probability 
\begin{equation*}
\Bbb{P}\left( \mathbf{I}_{t+1}=\mathbf{i}\mid \mathbf{I}_{t}=\mathbf{i}%
\right) =1-\sum_{\alpha ,\beta :\beta \neq \alpha }\frac{i_{\alpha }}{N}%
p_{\beta }\left( \frac{\mathbf{i}}{N}\right) =\sum_{\alpha }\frac{i_{\alpha }%
}{N}p_{\alpha }\left( \frac{\mathbf{i}}{N}\right) ,
\end{equation*}
completing the characterization of the $K-$allele Moran model. The
probability that in a one-step transition, the size of allele $A_{\alpha }$
population shrinks of one unit is: 
\begin{equation*}
\sum_{\beta \neq \alpha }P\left( \mathbf{i},\mathbf{i}_{\alpha ,\beta
}^{\prime }\right) =\frac{i_{\alpha }}{N}\left( 1-p_{\alpha }\left( \frac{%
\mathbf{i}}{N}\right) \right) .
\end{equation*}
The probability that in a one-step transition, the size of allele $A_{\beta
} $ population undergoes a one unit growth is: 
\begin{equation*}
\sum_{\alpha \neq \beta }P\left( \mathbf{i},\mathbf{i}_{\alpha ,\beta
}^{\prime }\right) =\left( 1-\frac{i_{\beta }}{N}\right) p_{\beta }\left( 
\frac{\mathbf{i}}{N}\right) .
\end{equation*}
As a nearest-neighbor random walk model, the Moran model has a much simpler
transition matrix $P$ of the Jacobi type. The equilibrium measure of the
chain again is: 
\begin{equation}
\pi _{eq}:=\sum_{l=1}^{K}\pi _{l}\left( \mathbf{i}_{0}\right) \delta _{%
\mathbf{i}_{l}^{*}},  \label{46}
\end{equation}
where $\mathbf{\pi }_{l}$ again solves the Dirichlet problem (\ref{21a}) but
with this new simpler Jacobi $P$.\newline

In what follows, we assume the special one-parameter selection model leading
to: 
\begin{equation*}
p_{\beta }\left( \frac{\mathbf{i}}{N}\right) =\frac{i_{\beta }}{N}\frac{%
1+sa_{\beta }}{1+sa}.
\end{equation*}
\newline

\textbf{Mean fitness.} Let us compute the LST of $\sum_{k}a_{k}I_{k}^{\prime
}$ in the context of a Moran model. We get the factorized form: 
\begin{equation*}
\Bbb{E}_{\mathbf{i}}\left( e^{-\lambda \sum_{k}a_{k}I_{k}^{\prime }}\right)
=\sum_{\alpha ,\beta :\alpha \neq \beta }e^{-\lambda \sum_{k}a_{k}i_{\alpha
,\beta }^{\prime }\left( k\right) }P\left( \mathbf{i},\mathbf{i}_{\alpha
,\beta }^{\prime }\right) +e^{-\lambda \sum_{k}a_{k}i_{k}}\sum_{\beta }\frac{%
i_{\beta }}{N}p_{\beta }
\end{equation*}
\begin{equation*}
=e^{-\lambda \sum_{k}a_{k}i_{k}}\left( \sum_{\alpha ,\beta :\alpha \neq
\beta }e^{-\lambda \sum_{k}a_{k}\mathbf{d}_{\alpha ,\beta }\left( k\right)
}P\left( \mathbf{i},\mathbf{i}_{\alpha ,\beta }^{\prime }\right)
+\sum_{\beta }\frac{i_{\beta }}{N}p_{\beta }\right)
\end{equation*}
\begin{equation*}
=e^{-\lambda \sum_{k}a_{k}i_{k}}\left( \sum_{\alpha ,\beta :\alpha \neq
\beta }e^{-\lambda \left( a_{\beta }-a_{\alpha }\right) _{{}}}\frac{%
i_{\alpha }}{N}p_{\beta }+\sum_{\beta }\frac{i_{\beta }}{N}p_{\beta }\right)
\end{equation*}
\begin{equation*}
=e^{-\lambda \sum_{k}a_{k}i_{k}}\left( \sum_{\beta }e^{-\lambda a_{\beta
}}p_{\beta }\sum_{\alpha \neq \beta }\frac{i_{\alpha }}{N}e^{\lambda
a_{\alpha }}+\sum_{\beta }\frac{i_{\beta }}{N}p_{\beta }\right)
\end{equation*}
\begin{equation*}
=e^{-\lambda \sum_{k}a_{k}i_{k}}\left( \sum_{\beta }e^{-\lambda a_{\beta
}}p_{\beta }\left( \sum_{\alpha }\frac{i_{\alpha }}{N}e^{\lambda a_{\alpha
}}-\frac{i_{\beta }}{N}e^{\lambda a_{\beta }}\right) +\sum_{\beta }\frac{%
i_{\beta }}{N}p_{\beta }\right)
\end{equation*}
\begin{equation*}
=\left( e^{-\lambda \sum_{k}a_{k}i_{k}}\right) \left( \sum_{\alpha }\frac{%
i_{\alpha }}{N}e^{\lambda a_{\alpha }}\right) \left( \sum_{\beta
}e^{-\lambda a_{\beta }}p_{\beta }\right) .
\end{equation*}
Recalling $\Delta w_{\mathbf{I}^{\prime }}\left( \frac{\mathbf{i}}{N}\right)
=s\sum_{k=1}^{K}\left( \frac{I_{k}^{\prime }}{N}-\frac{i_{k}}{N}\right)
a_{k},$ this leads in particular to (compare with (\ref{22a})): 
\begin{equation}
\Bbb{E}_{\mathbf{i}}\left( \Delta w_{\mathbf{I}^{\prime }}\right) =\frac{s}{N%
}\left( \Bbb{E}_{\mathbf{i}}\left( \sum_{k=1}^{K}I_{k}^{\prime }a_{k}\right)
-\sum_{k=1}^{K}i_{k}a_{k}\right) =\frac{s}{N}\left( \sum_{\beta }a_{\beta
}\left( p_{\beta }-\frac{i_{\beta }}{N}\right) \right)  \label{47}
\end{equation}
\begin{equation*}
=\frac{s^{2}}{N\left( 1+sa\right) }\left( \sum_{\beta }\frac{i_{\beta }}{N}%
a_{\beta }^{2}-a^{2}\right) >0.
\end{equation*}
The variance of $\Delta w_{\mathbf{I}^{\prime }}\left( \frac{\mathbf{i}}{N}%
\right) $ could easily be computed. Using the above result, we indeed get: 
\begin{equation*}
\Bbb{E}_{\mathbf{i}}\left( e^{-\sum_{l}\lambda _{l}I_{l}^{\prime }}\right)
=e^{-\sum_{l}\lambda _{l}i_{l}}\sum_{\alpha }\frac{i_{\alpha }}{N}e^{\lambda
_{\alpha }}\sum_{\beta }e^{-\lambda _{\beta }}p_{\beta },
\end{equation*}
giving the joint LST of $\mathbf{I}^{\prime }$ given $\mathbf{I}=\mathbf{i}$%
. Putting $\lambda _{l}=0$ if $l\neq k$, the $k^{th}-$marginal reads$:$%
\begin{equation*}
\Bbb{E}_{\mathbf{i}}\left( e^{-\lambda _{k}I_{k}^{\prime }}\right)
=e^{-\lambda _{k}i_{k}}\left( 1-\frac{i_{k}}{N}+e^{\lambda _{k}}\frac{i_{k}}{%
N}\right) \left( 1-p_{k}+e^{-\lambda _{k}}p_{k}\right)
\end{equation*}
which is of the random walk type. Indeed, we get: $\Bbb{P}_{\mathbf{i}%
}\left( I_{k}^{\prime }=i_{k}^{\prime }\right) =0$ if $i_{k}^{\prime }\neq
i_{k}\pm 1$ or $i_{k}^{\prime }\neq i_{k}$ and 
\begin{equation*}
\Bbb{P}_{\mathbf{i}}\left( I_{k}^{\prime }=i_{k}\right) =\left( 1-\frac{i_{k}%
}{N}\right) \left( 1-p_{k}\right) +\frac{i_{k}}{N}p_{k}
\end{equation*}
\begin{equation*}
\Bbb{P}_{\mathbf{i}}\left( I_{k}^{\prime }=i_{k}+1\right) =\left( 1-\frac{%
i_{k}}{N}\right) p_{k}\text{ ; }\Bbb{P}_{\mathbf{i}}\left( I_{k}^{\prime
}=i_{k}-1\right) =\frac{i_{k}}{N}\left( 1-p_{k}\right) .
\end{equation*}
In the special one-parameter case, we have 
\begin{equation*}
\Bbb{E}_{\mathbf{i}}\left( I_{k}^{\prime }\right) =i_{k}+\left( p_{k}-\frac{%
i_{k}}{N}\right) =i_{k}+\frac{i_{k}}{N}\left( \frac{1+sa_{k}}{1+sa}-1\right)
.
\end{equation*}
Using previously introduced notations, this gives a first moment estimator
of $s$ explaining best $\left( \mathbf{i;}i_{k,m}^{\prime },m=1,...,n\right) 
$ as: 
\begin{equation}
s_{k}^{*}=\frac{N\left( \left\langle i_{k}^{\prime }\right\rangle _{\mathbf{i%
}}-i_{k}\right) }{i_{k}\left( a_{k}-a\right) }.  \label{48}
\end{equation}

\section{Concluding Remarks}

In this Note our concern has been to introduce the general formalism of
evolutionary genetics dynamics under fitness, in both the deterministic and
stochastic setups, and chiefly in discrete-time. In the stochastic version
of the problem, both the Wright-Fisher and the Moran models were considered.
In the process, we revisited the various facets of the famous Fisher theorem
of natural selection in both the deterministic and stochastic formulations.
For the sake of simplicity of the exposition, we limited ourselves to a
simplified one-parameter model where the sole selection parameter is
unknown. Using these preliminary results and facts, we discussed the
estimation problems of the selection parameter based on a single-generation
frequency distribution shift under both deterministic and stochastic
evolutionary dynamics. To the best of the author's knowledge, this
particular way to address the estimation problem is new. It was stressed
that in our models, there were no mutation effects included. We plan to
include these effects in a forthcoming work. When mutations are present, the
situation changes drastically. Firstly, in the deterministic formulation,
the replicator dynamics combining fitness and mutations no longer is
gradient-like in general. Still, an internal equilibrium point can exist
were fitnesses to be multiplicative or would the mutation rates satisfy a
house of cards condition \cite{K1}. Secondly, in the stochastic formulation,
the Markov chains under study (either Wright-Fisher or Moran) are ergodic,
now with an invariant measure which is independent of the initial condition.
This also changes the picture drastically.\newline

\textbf{Acknowledgment.} To a large extent, this work was triggered by the
talk given by Professor Warren J. Ewens at the CIRM meeting in Marseilles,
on May, $29$, $2009$; although the topics dealt here with are slightly
different in spirit. Therefore, it owes much, if not all, to him. I take
this opportunity to thank the organizers, Etienne Pardoux and Amaury
Lambert, on behalf of the ANR MAEV headed by Sylvie M\'{e}l\'{e}ard, for
giving me the chance to attend this Conference.

\end{document}